# Intelligent Service Selection in a Multi-dimensional Environment of Cloud Providers for IoT Stream Data through Cloudlets

Omid Halimi Milani, S. Ahmad Motamedi and Saeed Sharifian

Dep. of Electrical Engineering, Amirkabir University of Technology, 15914, Tehran, Iran

{omidhalimimilani, motamedi, sharifian_s}@aut.ac.ir

**Abstract**

   *The expansion of the Internet of Things' (IoT) services and a huge amount of data generated by different sensors, signify the importance of cloud computing services like Storage as a Service more than ever. IoT traffic imposes such extra constraints on the cloud storage service as sensor data preprocessing capability and load-balancing between data centers and servers in each data center. Also, service allocation should be allegiant to the Quality of Service (QoS). The hybrid MWG algorithm is proposed in this work, which considers different objectives such as energy, processing time, transmission time, and load balancing in both Fog and Cloud Layer which were not addressed all together. The MATLAB script is used to simulate and implement our algorithms, and services of different servers, e.g. Amazon, Dropbox, Google Drive, etc. are considered. The MWG has 7%, 13%, and 25% improvement in comparison with MOWCA, KGA, and NSGAII in metric of spacing, respectively. Moreover, the MWG has 4%, 4.7%, and 7.3% optimization in metric of quality in comparison to MOWCA, KGA, and NSGAII, respectively. The new hybrid algorithm, MWG, yielded to not only consider three objectives in service selection as but also improve the performance in comparison to the works that considered one, or two objective(s). The overall optimization shows that the MWG algorithm has 7.8%, 17%, and 21.6% better performance in comparison with MOWCA, KGA, and NSGAII in the obtained best result by considering different objectives, respectively.*

## 1. Introduction

The Internet of Things (IoT) and their potential needs have enhanced the importance of cloud computing. The number of cloud-based servers that provide the proper structure for the manifestation of IoT is in an expanding manner. Plus, there are a large number of users and sensors that produce a high volume of data [1]. The proper usage of cloud services based on

QoS criteria is essential. Furthermore, sensors have a limited battery, and should work for a long time without a need to recharge. Distributed clouds can be useful for processing IoT sensors' data [2].

One of the issues that must be considered in modern systems is energy consumption, which should be considered on both sides, users, and servers. Furthermore, optimal energy consumption can have a significant impact on reducing air pollution. According to data released in 2014, the American data centers had consumed 70 billion KW/h of Electrical energy. Therefore, there are a lot of researches focus on energy consumption [3-5]. Large companies such as Amazon, Google, Salesforce, Microsoft, and IBM have begun to set up a new data center to host Internet applications and data processing centers. The current service providers give users the option of using services based on their demands and pay due to their usage (Pay as you go) [6]. The necessity to maintain data highlighted the usage of storage services. The high number of data may cause [7]. To address overload in one service, load balancing among different services was proposed in the literature [8, 9].

IoT devices include many examples of sensors in which they have different applications in smart homes, healthcare, transportation, building and cities. Storing data in cloud services is cost-effective, which is practical in addressing the high demand for IoT devices. However, cloud storage services face challenges such as energy consumption and load balancing among services [9, 10]. Existence of different service providers make it difficult to choose the proper service. Furthermore, the network has an arbitrary topology, and selecting an appropriate service among the available services of different service providers becomes another issue which should be considered. Another commonly discussed issue in cloud computing is the processing time. The needed time to process various data is based on data type, size and rates [11, 12]. There are several works in the area of service allocation.

To the best of our knowledge, there is no work with capabilities of minimizing energy consumption, reducing service completion time, in addition to addressing load balancing among services. Considering different services from a variety of service providers are also addressed in the present work. A hybrid modified MOWCA and GWO algorithm have proposed in the present study. The modified algorithm can explore better, as a result, better solution set can be proposed. Two layers of Fog and Cloud will be managed simultaneously to have comprehensive solution. Finally, structured technique for analyzing complex decisions in addition to different methods is used to evaluate the proposed algorithm.

The remainder of this paper organized as follows: in Section 2, a brief discussion about recent articles in this area are presented. Architecture, algorithm, mathematical models are presented in Section 3. In Section 4, the proposed method is evaluated and compared to other works. In Section 5, the paper is concluded.

## 2. Related Work

Cloud computing and IoT are used in a variety of fields, including medical engineering, and social media [13]. Variety of architectures and algorithms are proposed in this area to fulfil users' demands efficiently.

### 2.1. Architectures in Service Allocation

Some Cloud based architecture are centralized and the centralized architecture makes a latency in the processing and storing data. The latency is because of the great distance between IoT devices and cloud providers. Therefore, the Fog architecture can reduce latency. Managing energy consumption is presented in ECIoT architecture and leads to an optimal solution [14]. The idea of Fog computing is introduced in 2012 by researchers of the Cisco company with a new view to the set of all networks (including 3 G / 4 G / LTE / 5 G) and everything (all smart objects, Internet of Everything) with a hierarchical structure. By using the architecture expressed in Fog-cloud, the energy optimization of both cloud and Fog layers can be managed simultaneously to contribute the IoT development goal [15]. The architecture of Fog computing involves three main layers; IoT devices as the first layer, the fog layer as the second layer and finally, the cloud layer [5, 10].

Cloudlet is the project presented by a group of researchers from Mellon University[16]. The goal of using cloudlet is to have available resources near the IoT devices. Data compromise risk is reduced by having an edge computing. Cloudlets have more capacity for processing and battery power in comparison to other fog nodes.

### 2.2. Algorithms in Service Allocation

Researches have indicated that service allocation is NP-Hard problem [17]. A bunch of studies are concentrated on load balancing. Researchers are using definitive and innovative methods in a fair distribution of loads among services. The primary objective of load balancing is preventing overload in services. In [18], a random algorithm is used for outsourcing data to different services in which the physical distance is a criterion for selecting a service. In some cases, the service failure occurring due to lack of attention to the load balancing between services [19]. In [20], using the honey bee algorithm and introducing a new algorithm named honey bee behavior inspired load balancing (HBB-LB) are addressing equitable load distribution between virtual machines and maximizing processing speed. Round robin is also used for equitable load distribution. Broberg et al. [21] had chosen low-cost cloud storage services with Meta CDN method. They find and optimize service selection based on its cost, the process does not encompass an equitable load distribution between storage services [22, 23]. In [24, 25], a kind of NSGAII is used to address load balancing as the main criteria.

Some researches are performed to reduce completion time. Services have different processing speeds [26]. In [27], a model for load balancing on the Internet is presented, which aims to reduce the overall processing time for different tasks. In the other work, both the firefly algorithm and particle swarm optimization are made to balance the load of the entire system and reduce the makespan as well [28]. In [29], the MOGWO method, which is one of the Multi-objective Algorithms is used to distribute data equitably among virtual machines and also to reduce the working time. In [30], by using the GWO algorithm, the time of the makespan is reduced, and other determinants of QoS are not considered. In another research, the combined GWO and Cuckoo search algorithm is used to minimize the makespan and to reduce the needed services [31]; however, other aspects of service allocation are not considered. In [32], the integer programming method is used to reduce the cost of storing data and data retrieval time in a multi cloud provider environment. In [33], the particle swarm optimization (PSO) is used to minimize the cost of sending and processing of different tasks. The game theory is used for task scheduling in [34] for designing the mathematical model. The algorithm is proposed to deal with big data and to consider energy consumption. However, the available bandwidth as one of the factors in transmitting data and also energy consumption are not considered. In [35], an adaptive mode for assigning different services for tasks by considering the time is used. In [36], the NSGAII algorithm is used to achieve the minimum energy and makespan for service allocation in which load balancing is not considered. In [37], in addition to minimizing Energy consumption, the GWO and BAT algorithms are used to distribute the work equitably among services. In another research, energy and cost are optimized in service allocation [38]. In [39], the NSGAII algorithm is used to achieve the minimum energy and makespan for service allocation in which load balancing is not considered. The Pareto's results as one of important factors in Multi-objective problems are improved. Both Energy consumption and Load Balancing between solutions are considered while satisfying the spacing metric[40]

## 3. Proposed Algorithm for Fog-Cloud Architecture

The system model consists of different parts as illustrated in Fig. 1. The storing data as fast as possible by consuming less energy is critical. The first Cloudlets are assumed as the first sub layer in the presented model. Neighbor Cloudlets are assumed as a second sub layer in model. To accomplish the process of data as quickly as possible, the distribution of data among different Cloudlets is adopted to avoid overload. Using the distributed model is not only limited failure in service allocation, but also reducing energy consumption. The reliability of the system is reduced by rejecting request from different users. Furthermore, rejecting request destroys the Service Level Agreement. Distribution of data among different services increases the availability of services due to more available free space. The availability of different services is one of the QoS factors. So, the distribution of data among different services satisfies the QoS. By distributing data among services, failure in one service is not lead to loss of all data[5].

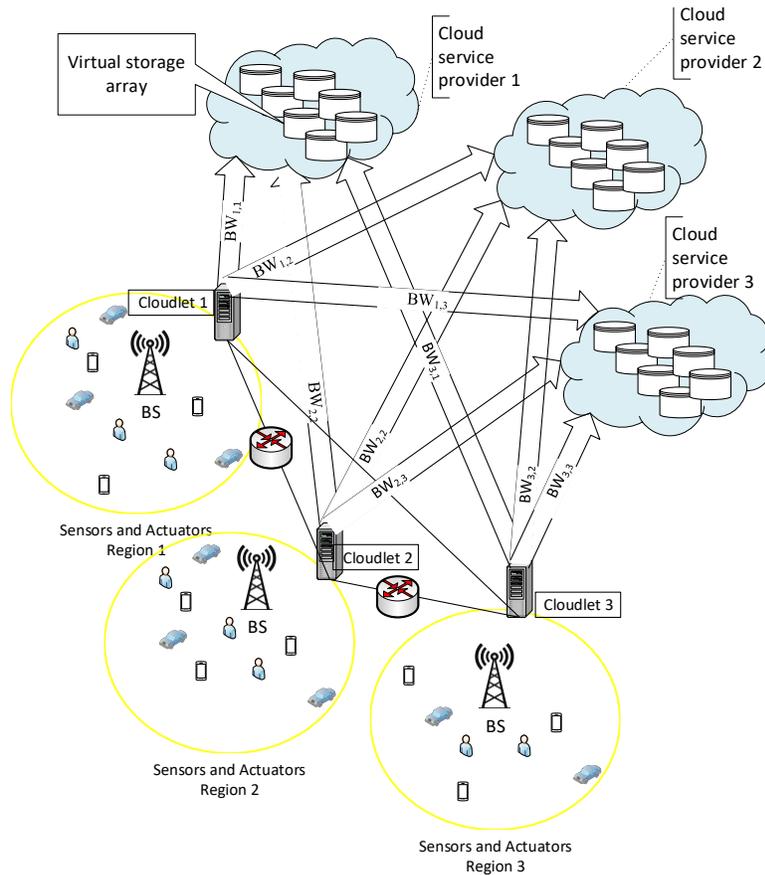

**Fig. 1.** Proposed architecture from sensors through cloudlet to cloud providers for storing IoT data.

Each Cloudlet manages a batch of data to be processed in or out of that particular Cloudlet. The proposed algorithm run in Cloudlet to find appropriate services in and out of Cloudlet in both Layers. The storage capacity and CPU are both considered in service allocation.

The process of outsourcing data is depicted in Fig. 2. For instance, a batch of data is received by the cloudlet 1 from its proximity sensors. Then, data are outsourced to proximity Cloudlets based on the status of network and services. Finally, data are transferred to the appropriate storage services based on the decision which are concluded in the first Cloudlet.

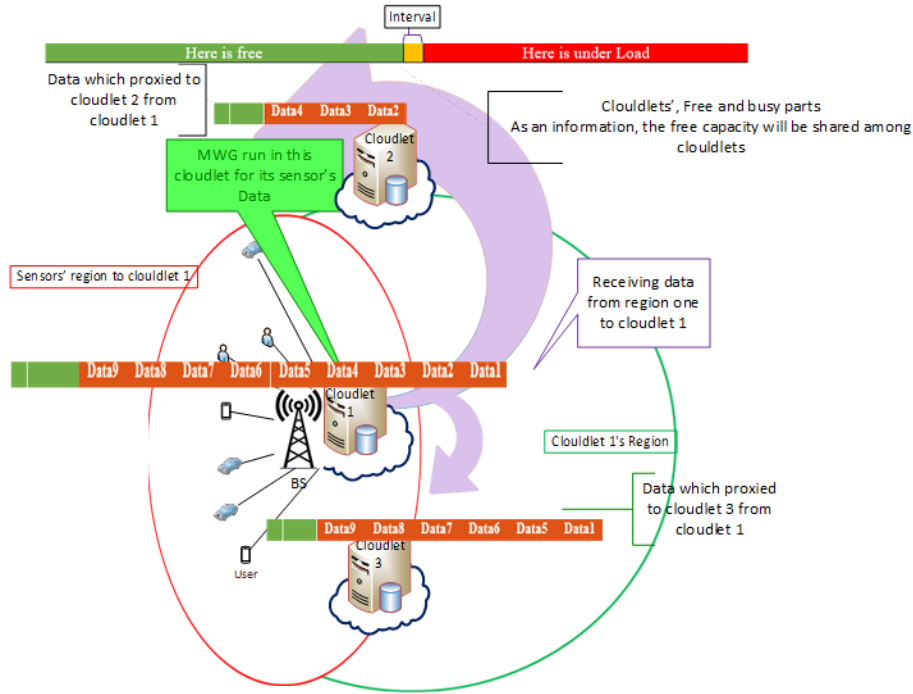

**Fig. 2.** Data allocation to different Cloudlets

## 3.1 Mathematics and Cost Functions

In this section, the mathematical model is introduced. In Table 1, the description of parameters is listed.

**Table 1.** Symbols and notations of objects' functions

| Symbol | Description |
|---|---|
| $E^{idle}$ | Energy consumption in an idle mode |
| $E^{busy}$ | Energy consumption in a busy mode |
| D | Data volume |
| m | Number of available services in Fog Layer |
| n | Number of data |
| $f_j$ | An active indicator of Cloudlet$_j$ |
| $f_{ij}$ | Allocate indicator of ith data to cloudlet$_j$ |
| $E_{tot}$ | Total energy consumption in each solution |
| $Cn_j$ | Number of CPU |
| $T_c$ | Time criterion for processing in the Fog layer |
| $capT_i/capC_i$ | The used percentage of Cloudlets. |
| $\overline{capT/capC}$ | The average usage ratio of Cloudlets. |
| $LB_c$ | Load-balancing criterion among services in Fog Layer. |
| $T_s$ | Consumed time criterion in transmitting to Cloud layer. |
| k | Number of storage services |
| $BW_{jz}$ | Bandwidth between j th cloudlet and z th cloud. |
| $LB_s$ | Load-balancing criterion among services in Cloud Layer. |
| P | Total number of populations in Multi-objective algorithms |
| $p_i$ | The ith solution in P. |
| $D_i$ | The ith data's volume |
| $C_a$ | The ath cloudlet's service |
| $cp_j$ | The capacity of the j's service. |
| $S_d$ | The dth cloud storage's service |

### 3.1.1 Mathematical Model for Energy Consumption

Servers consume energy in the idle and busy mode. According to [41], the energy consumption of servers can be represented by a linear relationship within the energy consumption and CPU utilization. Energy consumption can be calculated according to the proportion of service usages [42]. The total energy consumption in allocating services to a batch of data is $E_{tot}$.

$$E_{tot} = \sum_{j=1}^{m} \left[ f_j \times \left( \left( E_j^{busy} - E_j^{idle} \right) \times \frac{\sum_{i=1}^{n}\left( f_{ij} \times D_i \right)}{cp_j} + E_j^{idle} \right) \right] \quad (1)$$

$$j = 1, ..., m$$

$$\text{s.t.} \sum_{i=1}^{n}\left( f_{ij} \times D_i \right) \leq cp_j$$

The used energy in an idle and fully loaded modes are $E_{idle}$ and $E_{busy}$, respectively. The indicator D is used for demonstrating data size. In (1), m shows the number of CPU and n illustrate the number of data. $f_j$ has two values, 1 when the service is utilized and 0 when the service is not used. $f_{ij}$ has two values and it can be zero or one, it is 1 when data i is outsourced to service j.

### 3.1.2 Mathematical Model for Processing Time

Processing time has a direct relation to the size of data, and reverse relation to the number of CPU in one particular service according to (2). $T_c$ illustrates the total processing time for a batch of data

$$T_c = \max_j \left\{ \frac{\sum_{i=1}^{n} f_{ij} \times D_i}{Cn_j} \right\}; \; j = 1, 2, ..., m \qquad (2)$$

$f_{ij}$ has two values and it can be zero or one, it is 1 when data i is outsourced to service j.

### 3.1.3 Mathematical Model for Load Balancing in the Fog Layer

As mentioned, load balancing will increase the availability of services. The (3) address the load balancing in Fog-Layer [24].

$$LB_c = \sqrt{\frac{1}{n} \sum_{i=1}^{n} (capT_i/capC_i - \overline{capT/capC})^2} \qquad (3)$$

### 3.1.4 Mathematical model for transmission time

The data are outsourced to the appropriate storage service after being pre-processed. Eq. 6 shows the total transmission time for a batch of data.

$$T_s = \max_j \left\{ \sum_{i=1}^{n} \sum_{z=1}^{k} \frac{x_{iz} \times D_i}{BW_{jz}} \right\}; \; j = 1, 2, ..., m \qquad (6)$$

$BW_{jz}$ is an item of $BW_{c,s}$ matrix which shows the available bandwidth between Cloudlets and service providers. The rows of the matrix represent different Cloudlets, and the matrix's columns represent different storage service providers. z represents different storage services. $x_{iz}$ can be either zero or one based on storage services selection.

### 3.1.5 Mathematical Model for Load Balancing in the Cloud Layer

There are different services suggested from a variety of service providers. Eq. 7 performs balancing among various services.

$$LB_s = \sqrt{\frac{1}{n} \sum_{i=1}^{n} (capT_{si}/capS_i - \overline{capT_s/capS})^2} \qquad (7)$$

$LB_s$ as a factor should be minimized to have a reliable system. Different equations are addressed and all of which are important in service selection, as a result, the Multi-objective algorithms is proposed.

## 3.2. Proposed Hybrid Algorithm

Fig. 5 shows the block diagram of the proposed algorithm. Monitoring services, and making the decision based on the gathered information are done in the first cloudlet in which the data are received.

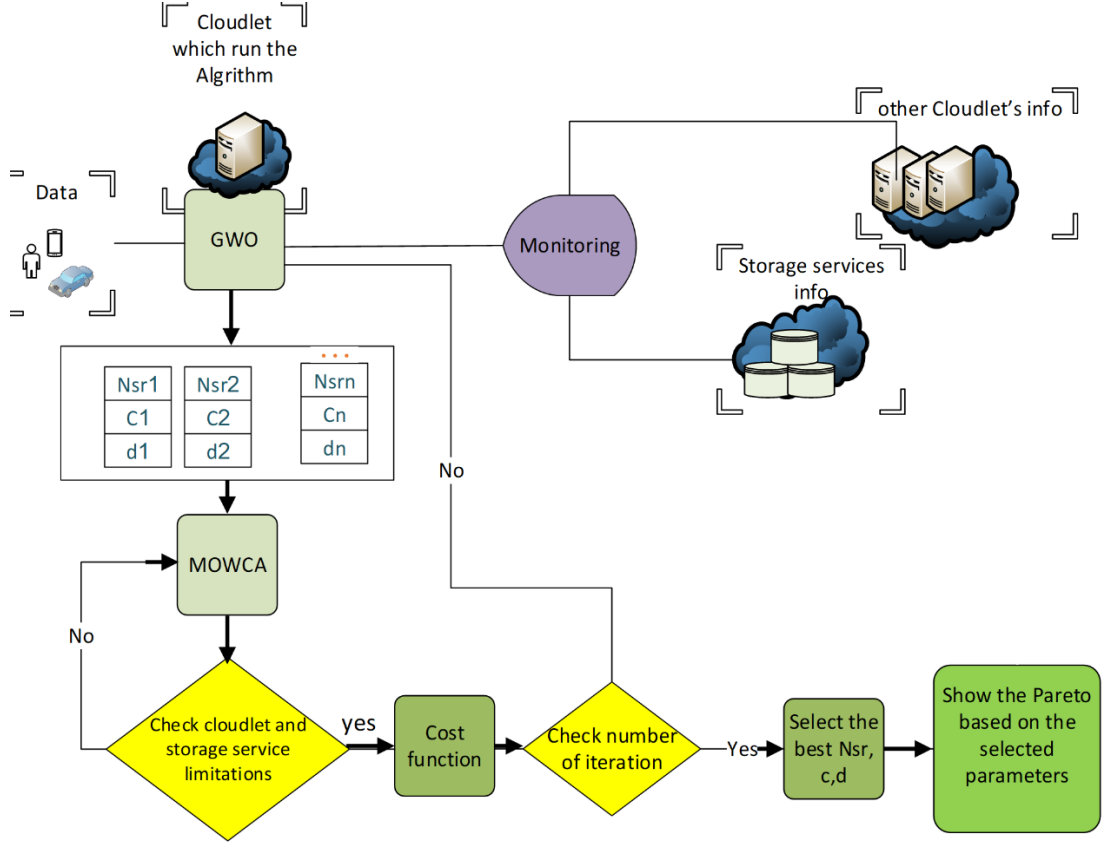

**Fig. 5.** Decision-Making block diagram

The Water Cycle Algorithm is adopted from nature. It is a population-based algorithm like a genetic algorithm [42]. Each member of the population represents as a 3 * n matrix (8).

$$p_i = \begin{Bmatrix} D_1 & D_2 & ... & D_n \\ C_a & C_b & ... & C_c \\ S_d & S_e & ... & S_f \end{Bmatrix} \quad (8)$$

For example, $D_1$ uses services of $C_a$ and $S_d$ in Fog and Cloud layer respectively. The cost of different layers are shown in Eq. 9 and Eq. 10.

$$Cost_{Fog\_layer} \propto E_{tot}, T_c, LB_c \quad (9)$$
$$Cost_{storage} \propto E_{tot}, T_s, LB_s \quad (10)$$

The goal is to minimize all five objects in the problem. No other solutions cannot dominate a solution in the Pareto. In mathematical terms, $p_1 \in P$ dominates $p_2 \in P$ if two conditions are satisfied [43].

$$\begin{aligned} Cost_i(p_1) \leq Cost_i(p_2) & \quad\quad \forall i \in \{E_{tot}, T_c, LB_c, T_s, LB_s\} \\ & \& \\ Cost_j(p_1) \prec Cost_j(p_2) & \quad\quad \exists j \in \{E_{tot}, T_c, LB_c, T_s, LB_s\} \end{aligned} \quad (11)$$

The number of solutions in the first iteration of the algorithm in resulted Pareto are low, but gradually, the number of non-dominated solutions are increased by moving the answers to the global minimum in the proposed algorithm. In the MOWCA, the solutions are governed by the sea and the rivers. The answers go toward Pareto, and Pareto is going to be updated. Finally, the answers in Pareto are categorized as a sea and rivers based on crowding distance [44].

### 3.2.1 Combination of GWO and MOWCA Algorithm (MWG)

Escaping from the local minimum is important and for this reason, determining appropriate $d_{max}$ is essential. Furthermore, G as a coefficient is used in the process of generating new solutions and it is crucial for a better exploration of solution space.

The GWO algorithm is one of the fastest optimization methods [45]. The MWG algorithm is the modified combined algorithm which optimize the hyper parameter of multi-objective algorithm and provide better exploration in the solution space in comparison to the naïve ones. The mathematical symbols which are used for GWO are shown in Table 2. At the beginning, a population with 10 members are randomly generated in GWO algorithm. Each member of this population has three features based on three hyper parameters of MOWCA. MOWCA run and optimized the objects based on the initialized values. Then, the MWG determines the crowding distance and dominate solutions. Crowding-distance is calculated for all non-dominant solutions. The main objective of using crowding distance is to increase the diversity of solutions. In the rest of the paper, the mathematics of these algorithms are discussed.

**Table 2.** Mathematical symbols of a proposed algorithm

| Symbol | Description |
|---|---|
| $p_{Stream}^j$ | Solution j which consider as a stream in the algorithm. |
| $p_{Sea}^j$ | Solution j which consider as a Sea in the algorithm. |
| $p_{River}^j$ | Solution j which consider as a River in the algorithm. |
| G | The coefficient for generating new solutions in MOWCA algorithm. |
| $d_{max}^i$ | The maximum acceptable distance between solutions and the Sea in MOWCA algorithm. |
| $\bar{X}$ | The position of a solution as a hunt in the GWO algorithm. |

| $\vec{X}_\alpha$ | The position of a solution as a $\alpha$ in the GWO algorithm. |
|---|---|
| $\vec{X}_\beta$ | The position of a solution as a $\beta$ in the GWO algorithm. |
| $\vec{X}_\delta$ | The position of a solution as a $\delta$ in the GWO algorithm. |
| $\vec{X}(t+1)$ | The new result which is calculated for Nsr, C, and $d_{max}$. |

In MOWCA, $p^j_{Stream}$ is one of the answers. Rand is a number between 0 and 1. In the exploration phase of the algorithm new positions for rivers and streams are proposed by Eq. 12-14.

$$p^{j+1}_{Stream} = p^j_{Stream} + rand \times G \times (p^j_{Sea} - p^j_{Stream}) \qquad (12)$$
$$p^{j+1}_{Stream} = p^j_{Stream} + rand \times G \times (p^j_{River} - p^j_{Stream}) \qquad (13)$$
$$p^{j+1}_{River} = p^j_{River} + rand \times G \times (p^j_{Sea} - p^j_{River}) \qquad (14)$$

Large value for $d_{max}$ prevents from convergence, and a smaller value of $d_{max}$ tends to trap in the vicinity of the sea and reduces the exploration. The value of $d_{max}$ controls the position of different solutions around the sea, and it decreases gradually based on Eq. 15.

$$d^{i+1}_{max} = d^i_{max} - \frac{d^i_{max}}{Max(itr)} \qquad (15)$$

Nsr, $d_{max}$, and G are optimized through the MWG algorithm based on the Pareto. The best hunt agent ($\vec{X}_\alpha$), the second best hunt agent ($\vec{X}_\beta$) and the third best hunt agent ($\vec{X}_\delta$) are considered as alpha, beta, and delta using Equations 16 and 17. And, X represents the position of a hunt.

$$\begin{aligned} \vec{D}_\alpha &= |\vec{F}\vec{X}_\alpha - \vec{X}| \\ \vec{D}_\beta &= |\vec{F}\vec{X}_\beta - \vec{X}| \\ \vec{D}_\delta &= |\vec{F}\vec{X}_\delta - \vec{X}| \end{aligned} \qquad (16)$$

$$\begin{aligned} \vec{X}_1 &= |\vec{X}_\alpha - \vec{A}(\vec{D}_\alpha)| \\ \vec{X}_2 &= |\vec{X}_\beta - \vec{A}(\vec{D}_\beta)| \\ \vec{X}_3 &= |\vec{X}_\delta - \vec{A}(\vec{D}_\delta)| \end{aligned} \qquad (17)$$

The vectors of A and F are calculated based on Eq. 18.

$$\begin{aligned} \vec{A} &= 2\vec{a}\vec{r}_1 - \vec{a} \\ \vec{F} &= 2\vec{r}_2 \end{aligned} \qquad (18)$$

The r1 and r2 are vectors in the range of [0,1]. Finally, the position of a hunt calculated based on Eq. 19.

$$\bar{X}(t+1) = \frac{\vec{X}_1(t) + \vec{X}_2(t) + \vec{X}_3(t)}{3} \tag{19}$$

To get close to the prey, the amount of $\vec{a}$ reduced from 2 to 0 gradually. All the values in the obtained Pareto are normalized based on their objectives. Then, the summation of all five objectives for each solution in the Pareto is calculated based on Eq. 20. The best answer choice is the one with minimum cost.

In the next iteration, for normalizing the achieved Pareto, the previous selected result as a best are considered. Unlike single objective problems which are concluded in one cost at the end of each iteration, there is a Pareto in multi-objective algorithms which involves different solutions at the end of each iteration.

(20)

Finally, the optimized values for $d_{max}$, Nsr, and G along with the Pareto are achieved.

## 4. Simulation

Number of CPU in a service plays an important role in the processing speed. For example, t2.nano has a one CPU, and t2.large has two CPUs., all of which are assumed in the simulation environment as a variety of available services [46]. The Cloud services have different throughputs, which lead to different transmission time. Table 3 illustrates the different cloud service providers' throughput. Different size of data are considered according to real IoT sensors' data for storing in three batches. For different scenarios, different batch sizes are considered. Also, 80% of services contemplated to be needed to fulfil the demand for batch of size 400 to validate the reliability of service allocation. The data size has a uniform distribution between 20 and 500 for each batch of data [47, 48].

**Table 3.** Different Cloud Service Providers' Throughput [50]

| Cloud Service Provider | Throughput (Mbps) |
|---|---|
| Amazon S3a | 1.349 |
| Box | 2.128 |
| Dropbox | 2.314 |
| OneDrive | 2.233 |
| Google Drive | 4.465 |
| SugarSync | 2.171 |
| Cloud Mine | 1.474 |
| Rackspace | 1.704 |

There is a relationship between the number of CPU and energy consumption in each service. According to the SPECpower benchmark, the maximum power consumption assumed 250 W for each CPU. To consider energy, 200 W in the idle mode and 300 W in the busy mode are assumed for each CPU [49].

### 4.1. Simulation Setups

It is assumed that there are 32 storage services in the Cloud layer and 32 processing services in Fog layer. There are Multi-objective algorithms which addressed energy and load-balancing in service selection in recent years such as NSGAII and the modified GA called KGA. In the KGA, k-means is used to have an elitist set of the population [24]. The average results of 10 separate run for each algorithm is calculated. The Pareto's answers are sorted and the mean of 10 separate Pareto is calculated for each objective. The results are reported in BOX diagram, that has quartet and means. The box diagram gives a better visual distribution of solutions for each objective in resulted Pareto. Numerical experiments are conducted using MATLAB on a Laptop with a Core i7 2.59-2.60 GHz CPU. The number of population for the first part of the MWG algorithm is 10, and for the second part of the MWG algorithm is 50.

The setting parameters of other algorithms in the simulation are listed in Table 4.

**Table 4.** Different algorithms and the setting Parameters

| Algorithms | Parameters | Settings |
|---|---|---|
| NSGAII | Population size(pop), Crossover probability, Mutation probability | 50,0.8,1 |
| KGA | Population size(pop), Crossover probability, Mutation probability, number of centroids | 50,0.8,0.1,4 |
| MOWCA | Population size(pop), Number of streams, the distance of sea(dmax) | 50,4,1 |

## 4.2. Evaluation

In Fig. 7 the distribution of different solutions' energy consumption in Pareto of various algorithms are shown.

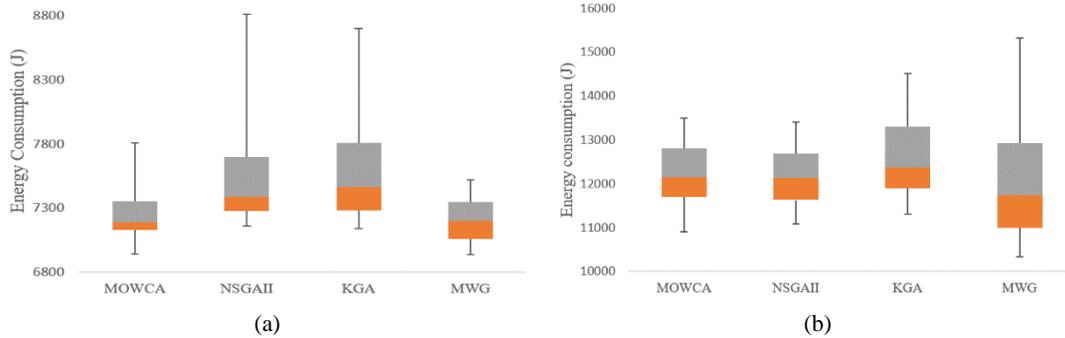

**Fig 7.** Comparison between algorithms due to the distribution of different solutions' energy consumption in the resulted Pareto after 100 iterations for (a) 100 Data and (b) 400 Data

In the MWG, rivers and sea lead other answers toward the optimum solution. The number of solutions in Pareto for MWG is more than the other two KGA, and NSGAII algorithms based on energy consumption criteria. Fig. 8 shows the distribution of different solutions' processing time in the Pareto for different algorithms. If the time plays a critical role, the MWG recommends some solutions with less processing time in comparison to other algorithms even for a specific case, like maximum demand scenario.

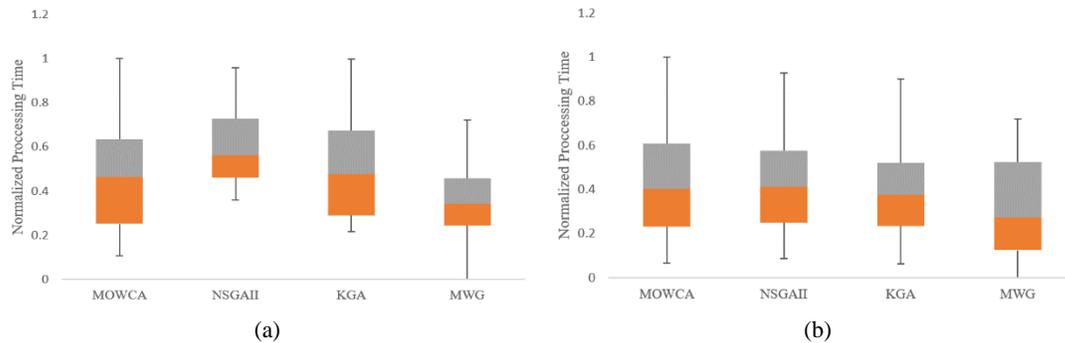

**Fig. 8.** Comparison between algorithms due to the distribution of different solutions' Processing time in the resulted Pareto after 100 iterations for (a) 100 Data and (b) 400 Data

Another factor that should be considered for the service allocation is the load balancing between the VMs in the cloudlet as well as services from different Cloudlets.

Fig. 9 shows the distribution of different solutions' load Balancing in the Pareto of various algorithms. Service may fail because of the overload. Thus, having a fair distribution of data among services increases the reliability as one of the critical factors in QoS. NSGAII and KGA use the method of mutation and crossover. These methods are good in escaping from Local minimum, they are random based methods in which the global minimum may be ignored as well.

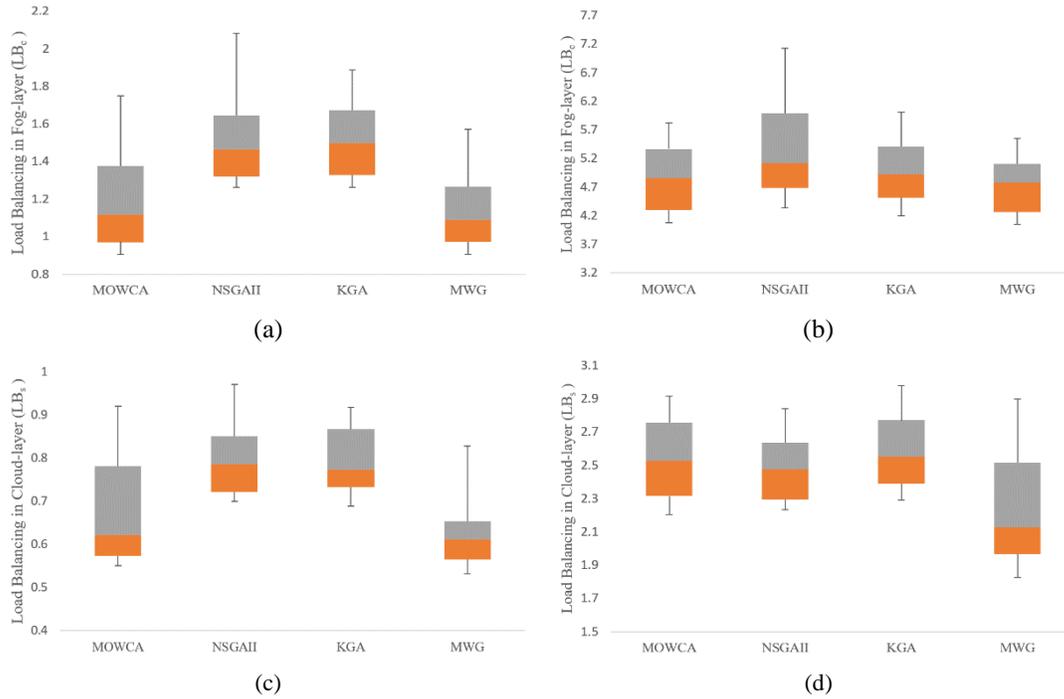

**Fig. 9.** Comparison between algorithms due to the distribution of different solutions' load balancing in the resulted Pareto after 100 iterations for (a) 100 Data in Fog Layer (b) 400 Data in Fog Layer (c) 100 Data in Cloud Layer, (d) 400 Data in Cloud Layer

On the other hand, many services like Dropbox, after processing data, outsource data to cloud storage services for storing. Service providers around the world provide a variety of services for users. The throughput of different service providers varies. Therefore, the needed time to outsource data to cloud services varies based on available Bandwidth. Fig. 10 shows the distribution of different solutions' transmission time in the Pareto of various algorithms. NSGAII and KGA have no leader in their pool of answer choices. The solutions in NSGAII are randomly generated based on crossover and mutation.

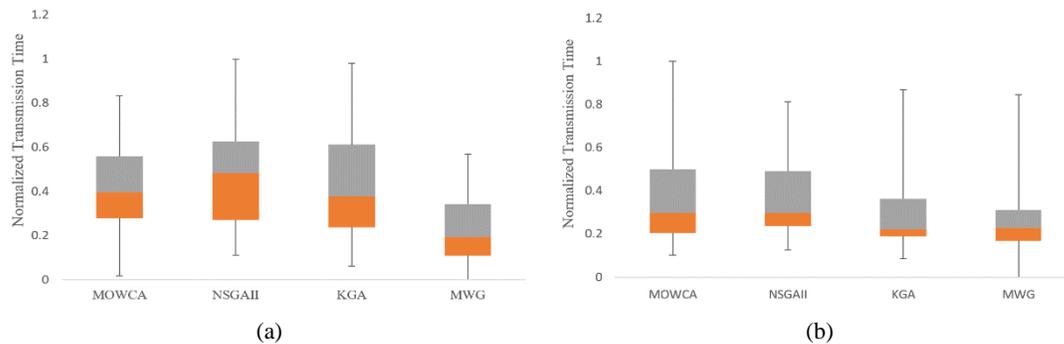

Fig. 10. Comparison between algorithms due to the distribution of different solutions' transmission time in the resulted Pareto after 100 iterations for (a) 100 Data and (b) 400 Data

Table. 5 shows the percentage of solutions in which the MWG algorithm has a better result in the Pareto in comparison to other algorithms. To have a comparison among other algorithms, colours are used. For instance, the green shows the second good algorithm, yellow shows the third-ranked among all algorithms, and the red is the fourth-ranked algorithm.

Table 5. Percentage of better results in the Pareto of MWG in comparison to other algorithms.

| Main Algorithm | Objectives | 100 data | | | 400 data | | |
|---|---|---|---|---|---|---|---|
| | | KGA | MOWCA | NSGAII | KGA | MOWCA | NSGAII |
| MWG | Energy | 34% | 4% | 40% | 36% | 24% | 26% |
| | Load Balancing Fog | 72% | 6% | 74% | 22% | 8% | 30% |
| | Load Balancing Cloud | 80% | 18% | 80% | 64% | 54% | 56% |
| | Processing Time | 18% | 10% | 56% | 6% | 8% | 16% |
| | Transmission Time | 14% | 38% | 26% | 14% | 4% | 26% |

## 4.3 Metric of Spacing

The statistic factor for comparing two different multi-objective algorithms is the regularity of answers in the Pareto. This factor is demonstrated with a metric of spacing. The Eq. 21 is used for determining regularity in different algorithm's Pareto.

$$SP = \sqrt{\frac{1}{n_{pf}-1}\sum_{i=1}^{n_{pf}}(d_i - \bar{d})^2} \qquad (21)$$

$n_{pf}$ shows the number of solutions in Pareto. $d_i$ shows the distance between every two answer choices in the Pareto. The average distance between every two answer choices in the Pareto is shown by $\bar{d}$. If $SP$ is smaller, the solutions in the Pareto are more regularly distributed. In Table 4, results for different algorithms are shown (rounded up). According to Table 6, the average metric of spacing for MWG has 7%, 13%, and 25% optimization in comparison to the average metric of spacing for MOWCA, KGA, and NSGAII respectively.

Table 6. Metric of Spacing for different algorithms.

| Data number | Iteration | KGA | | NSGAII | | MOWCA | | MWG | |
|---|---|---|---|---|---|---|---|---|---|
| | | Min S | Max S | Min S | Max S | Min S | Max S | Min S | Max S |
| 100 | | 130 | 134 | 165 | 189 | 102 | 112 | 114 | 139 |
| 200 | 100 | 237 | 244 | 257 | 259 | 232 | 248 | 210 | 215 |
| 400 | | 311 | 325 | 352 | 389 | 304 | 314 | 267 | 269 |

## 4.4 Metric of Quality

The quality of the Pareto shows the difference between the optimum and obtained results. Since the optimum solutions are not defined in NP-HARD problems, the new method is suggested in which the quality of Pareto among different algorithms can be compared.

In the Algorithm 1, all solutions are gathered from different algorithms' Pareto. Solutions are sorted based on the non-dominated solution. Then, solutions are sorted based on crowding-distance. Finally, an aggregated Pareto is achieved. The number of solutions from each algorithm's Pareto in the Aggregated Pareto shows the quality of different algorithms in comparison to each other. By considering all workloads, the MWG algorithm has 4%, 4.7%, and 7.3% advantages in metric of quality in comparison to MOWCA, KGA, and NSGAII respectively.

| **Algorithm 1:** Pareto Quality Function (Metric of Quality) |
|---|
| **Input:** Different Algorithms' solutions in their Pareto |
| **Output:** The percentage of each Algorithms' solutions in the new optimum Pareto |
| **Begin** |
| 1: Insert Pareto of algorithms |
| 2: pop <= Non Dominated Sorting(pop);[44] |
| 3: CD <= Calculate Crowding Distance(pop);[44] |
| 4: Aggregated Pareto = Sort Population(pop) based on CD |
| 5: $N_a$ = Number of Algorithms; |
| 6: $N_{sp}$ = Number of Solutions in Aggregated Pareto; |
| 7:     **For** k=1: $N_a$ |
| 8:         **For** j=1: Number of Solutions in $k^{th}$ Algorithm's Pareto |
| 9:             **If** Solution(j) is in Aggregated Pareto set |
| 10:                 N(k) = N(k) + 1; |
| 11:             **End if** |
| 12:         **End for** |
| 13:         Metric of Quality (k) = $\dfrac{N(k)}{N_{sp}}$ ; |
| 14:     **End for** |
| **End** |

The percentage of solutions from different algorithms' Pareto which are involved in the Aggregated Pareto are demonstrated in Fig.12.

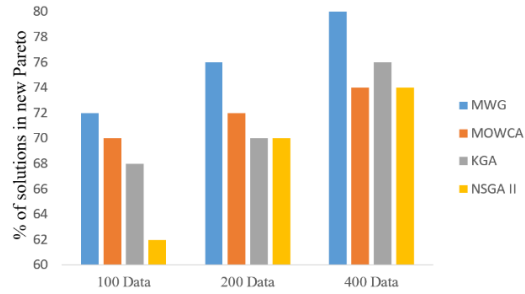

**Fig. 12.** Different algorithms' Pareto involvement in the Aggregated Pareto resulted in 100 iterations.

## 4.5 Evaluation of the optimum result

The answer choice with the minimum cost is considered as the best in each Pareto based on Analytic hierarchy process (AHP). The alternatives are solutions in the Pareto and criteria are the mentioned five objectives [50]. All objectives are considered to have the same importance. The best solution among the Pareto's solutions of each algorithm is determined based on AHP. Table 7 shows a comparison between different algorithms in each objective. For instance, the MWG has 1.37 percent improvement in energy consumption for 100 data.

Table 7. Comparison of algorithms' best solution based on AHP

| Main Algorithm | Objectives | 100 Data | | | 200 Data | | | 400 Data | | |
|---|---|---|---|---|---|---|---|---|---|---|
| | | MOWCA | KGA | NSGAII | MOWCA | KGA | NSGAII | MOWCA | KGA | NSGAII |
| MWG | Energy | 1.37 | 4.11 | 5.48 | 1.79 | 3.06 | 5.10 | 1.60 | 3.20 | 4 |
| | Load Balancing Fog | 4.95 | 30.69 | 33.66 | 6.57 | 14.14 | 15.15 | 5.13 | 15.38 | 20.51 |
| | Load Balancing Cloud | 6.12 | 30.01 | 33.71 | 10.08 | 24.11 | 26.28 | 16.68 | 26.05 | 30.23 |
| | Pros Time | 4.31 | 15.47 | 34.72 | 20.93 | 32.15 | 37.80 | 19.33 | 29.52 | 32.69 |
| | Transmission Time | 5.64 | 11.54 | 17.69 | 6.63 | 8.63 | 13.95 | 6.11 | 9.80 | 14.17 |

The overall optimization results show that MWG algorithm has 7.8%, 17%, and 21.6% better performance in comparison with MOWCA, KGA, and NSGAII based on obtained result from Pareto.

## 5. Conclusion

This paper presented a new hybrid Multi-objective algorithm for storage service selection in a collaborative and heterogeneous cloud and fog environment. Five objectives are considered altogether in the proposed work. The Mathematical models for energy, load-balancing, and time are proposed. Numerical experiments are conducted to compare the performances of the proposed MWG, MOWCA, KGA, and NSGAII. Testing results in a case study demonstrate the better performances of MWG in metric of spacing and metric of quality in comparison to

MOWCA, KGA, and NSGAII. Energy consumption is reduced, fair load distribution among services is achieved, and the time for processing and transmitting are optimized. The Pareto and the optimum result show the superiority of the proposed hybrid MWG algorithm in comparison to other algorithms in service selection. In future works, the influence of other parameters such as the node price will be added as an object to the mathematical model, which will be analyzed in more details. Besides, other practical applications will be considered such as latency, and user's demands and constraints. Furthermore, the algorithm with better performance in terms of Quality of Service can be proposed.